\documentclass[pageno]{jpaper}

\usepackage[normalem]{ulem}

\usepackage{multicol}
\usepackage[normalem]{ulem}
\usepackage[ruled,vlined,linesnumbered]{algorithm2e}
\usepackage[noend]{algpseudocode}
\usepackage{array}
\usepackage{amsmath}
\usepackage{bold-extra}
\usepackage{todonotes}
\usepackage{flushend}
\usepackage{enumitem}

\usepackage{graphicx}%
\usepackage{listings,lstautogobble}
\usepackage{xcolor}
\usepackage{multirow}
\newcolumntype{P}[1]{>{\centering\arraybackslash}p{#1}}
\newcolumntype{M}[1]{>{\centering\arraybackslash}m{#1}}

\usepackage{subfig}
\usepackage[title]{appendix}

\DeclareMathAlphabet{\mathcal}{OMS}{cmsy}{m}{n}

\SetKwProg{MyStruct}{struct}{ contains}{end}

\lstdefinestyle{CStyle}{
    backgroundcolor=\color[HTML]{f1f1f1},   
    commentstyle=\color[HTML]{000000}\small,
    keywordstyle=\color[HTML]{bd417b},
    basicstyle=\ttfamily\footnotesize,
    breakatwhitespace=false,         
    breaklines=true,                 
    captionpos=b,                    
    tabsize=0,
    showtabs=true,
    showspaces=false,
    language=C
    }

\newcommand{\name}{{\sf LEASH }}
\newcommand{\nameA}{{\sf LEASH}}

\begin{document}

\title{LEASH: Enhancing Micro-architectural Attack Detection \\ with a Reactive Process Scheduler}
\author{Nikhilesh Singh, Chester Rebeiro\\
Indian Institute of Technology, Madras\\
\{\em nik,chester\}@cse.iitm.ac.in}
\date{}
\maketitle
\thispagestyle{empty}

\begin{abstract}
Micro-architectural attacks use information leaked through shared  resources to break hardware-enforced isolation. These attacks have been used to steal private information ranging from cryptographic keys to privileged Operating System (OS) data in devices ranging from mobile phones to cloud   servers.
Most existing software countermeasures either have unacceptable  overheads or considerable false positives. Further, they are designed for specific attacks and cannot readily adapt to new variants.  

In this paper, we propose a framework called \nameA, which works from the OS scheduler to stymie micro-architectural attacks with minimal overheads,  negligible impact of false positives, and  capable of handling  a wide range of attacks. \name~works by starving maliciously behaving threads at runtime, providing insufficient time and resources to carry out an attack. The CPU allocation for a falsely flagged thread found to be benign is  boosted to minimize overheads. 
To demonstrate the framework, we modify Linux's Completely Fair Scheduler with \name and evaluate it with seven micro-architectural attacks ranging from Meltdown and Rowhammer to a TLB covert channel. The runtime overheads are evaluated with a range of real-world applications and found to be less than 1\% on average.

\end{abstract}

\section{Introduction} \label{sec:intro}

For several years, computer systems designers have relied on traditional hardware mechanisms such as protection rings, segmentation, page tables, and enclaves, to isolate software entities executing in a CPU. These isolation mechanisms have been considerably weakened by a potent class of side-channel attacks known as micro-architectural attacks. 
A micro-architectural attack makes use of shared hardware resources to leak information across isolation boundaries. They have been used in a variety of applications, from creating covert channels~\cite{hu:92}, retrieving secret keys of  ciphers~\cite{Bernstein:2005:CacheAttack, percival:05}, reading Operating System data~\cite{Kocher:2018:Spectre,Lipp:2018:Meltdown},  fingerprinting websites~\cite{Shusterman:2019}, logging keystrokes~\cite{Ristenpart:2019:Heyyou}, reverse engineering Deep Learning algorithms~\cite{Hong:2018:DL}, and breaking Address Space Layout Randomization~\cite{Barresi:2015:CAIN, Gras:2017:ASLR}.
The last couple of years have seen these  attack vectors applied on a  range of devices from mobile phones~\cite{Lipp:2016:ARMageddon} to third-party cloud  platforms~\cite{Ristenpart:2019:Heyyou}. They have been used to leak secrets stored in Trusted Execution Environments like SGX~\cite{Brasser:2017:SGXattack} and Trustzone~\cite{Zhang:2018:Trustzone} and attack remote computers using webpages that host malicious Javascript~\cite{Oren:2015:jscacheattack} or Web-assembly~\cite{Genkin:2018:drive-bycache}.

\begin{figure}[!t]
\captionsetup[subfloat]{farskip=2pt,captionskip=1.5pt}
\subfloat[\name stymies micro-architectural attacks by detecting malicious behavior in programs and reducing its CPU-share, thereby reducing the leakage from the shared resource.]{
  \includegraphics[width=\columnwidth]{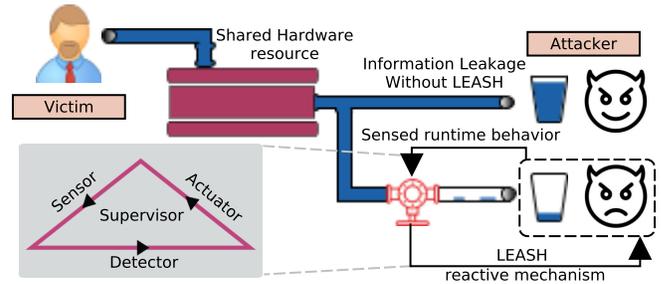} \label{fig:leash_analogy}}
  
\subfloat[\name detects malicious behavior by monitoring run time characteristics of all threads (${\tt t_1, t_2, t_3}$) in the system using HPCs.  Threads with a high threat index get less CPU time. Thread ${\tt t_1}$ has persistent malicious behavior and hence gets throttled. Thread ${\tt t_2}$ gets flagged for an epoch but recovers, while thread ${\tt t_3}$ remains unaffected.]{%
  \includegraphics[width=\columnwidth]{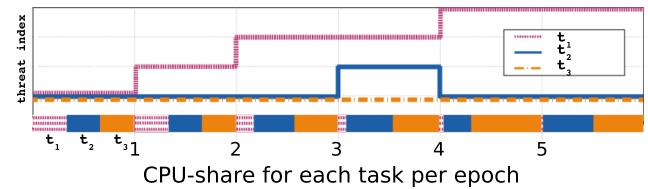}\label{fig:leash_highlevel}}
\vspace{-0.1cm}
\caption{A high-level overview of ~\nameA. }
\vspace{-0.2cm}
\end{figure}

In a typical micro-architectural attack, the attacker runs a program called the spy that contends with a victim program for shared hardware resources such as a common Cache Memory~\cite{ Aciimez:2010:instncache, Gruss:2015, Lipp:2016:ARMageddon, Liu:2015:LLC,percival:05}, Branch Prediction Unit (BPU)~\cite{Aciicmez:2007:BPU, Evtyushkin:2018:Branchscope}, Translation Lookaside Buffer (TLB)~\cite{Gras:2018:TLB}, or DRAM~\cite{Mutlu:2014:Rowhammer,Pessl:2016:DRAMA}. The contention affects the spy's execution time in a manner that correlates with the victim's execution. If the victim's execution pattern happens to depend on secret data, then, the correlation can be used to reveal it.

Most schemes to prevent micro-architectural attacks work either in the software  or are incorporated in the hardware. Hardware solutions include bypassing~\cite{Yan:2018:invisiSpec} or partitioning~\cite{intel:2015:CAT, Liu:2016:CATalyst, Wang:2006:RPcache} the shared resource, or randomizing accesses  to it~\cite{Qureshi:2018:Ceaser, Qureshi:2019:CS, Wang:2006:RPcache}. The preventive solutions in  software  attempt to have constant-time implementations~\cite{Yee:2009:Nacl} or inject noise in measurements to make the attack  difficult~\cite{hu:91:fuzzyTime,Martin:2012:Timewarp}.
While software solutions have huge overheads and may be environment specific,  hardware solutions are not applicable to legacy systems. 

An alternate approach that promises to solve these problems is to detect and thwart an attack in real time. The detection is typically done using  performance monitoring registers in the CPU called Hardware Performance Counters (HPCs)~\cite{Alam:2017, Aweke:2016:anvil,Briongos:2018:cacheShield,Chiapetta:2016:hpcdetection,Mushtaq:2018:NightsWatch,Zhang:2016:cloudRadar}. On detection, the malicious-behaving thread is terminated or rescheduled to a different CPU. A severe limitation of current detection schemes is the false positives due to which benign processes are falsely flagged malicious. Contemporary works~\cite{Alam:2017,Chiapetta:2016:hpcdetection,Mushtaq:2018:NightsWatch,Mushtaq:2020:whisper,Zhang:2016:cloudRadar} attempt to reduce false positives by deploying complex Machine Learning models. While this increases overheads, the impact of the false positive rate is still unacceptably high. This limitation is exacerbated by the fact that a micro-architectural attack occurs rarely compared to the number of benign programs executing. Thus, these detection techniques more often impact benign program execution than mitigate an attack.

In this paper, we present a micro-architectural attack detection countermeasure called ~\name that has all the advantages of detection countermeasures with almost no impact of false positives. ~\name makes use of the observation that a spy thread in a micro-architectural attack needs to contend with the victim for a shared resource. The success of the attack depends on the extent to which the spy can force this contention. If the spy gets insufficient time to execute on the CPU, then the information leakage is reduced, stymieing the  attack. Figure~(\ref{fig:leash_analogy}) provides an analogy.

The \name framework works from the Operating System (OS) scheduler. At every context switch, it uses the HPCs to quantify the malicious behavior of each thread in the system using a metric called {threat index}. It then allocates  CPU resources for the thread based on its {threat index}. A high value of {threat index} indicates that the process gets less CPU time. For instance, in Figure~(\ref{fig:leash_highlevel}) thread $\tt t_1$ has a {threat index} that continuously increases. The time slice it obtains for execution correspondingly reduces. If the thread were indeed a micro-architectural attack, the lesser CPU time would clamp the amount of information that is leaked from the victim, thus stymieing the attack. 

A significant advantage of~\name is in its handling of false detection. Suppose a benign thread temporarily exhibits malicious behavior, its CPU quota falls temporarily, and over time it regains its full CPU quota. Thread ${\tt t_2}$ in Figure~(\ref{fig:leash_highlevel}) demonstrates this recovery after the 4-th epoch. This is unlike contemporary works~\cite{Alam:2017,Chiapetta:2016:hpcdetection,Mushtaq:2018:NightsWatch,Mushtaq:2020:whisper,Zhang:2016:cloudRadar}, where a false detection has adverse irreversible effects. With \name, the false detection would only result in a small overhead. With the wide range of benchmarks we evaluated, the performance overheads were found to be less than 1\% on average.

\vspace{-0.3cm}
{\flushleft The important contributions of this paper are as follows:}
\begin{itemize}[wide,labelwidth=!,labelindent=1pt]
    \item \name is an OS scheduler based countermeasure for micro-architectural attacks that has low overheads and negligible impact of false positives. It can support a variety of micro-architectural attacks and can easily adapt to new variants.The framework is highly versatile and can be tuned to detect multiple different attacks simultaneously. It is designed
    to provide quick response, while still having features for sophisticated add-ons that can detect complex scenarios.
    \item We modify Linux's Completely Fair Scheduler (CFS) with \name and evaluate its effectiveness with seven micro-architectural attacks including  Prime+Probe on the L1 Data~\cite{Osvik:2006:cacheFlushing}, L1 Instruction~\cite{Aciimez:2010:instncache}, and LLC caches~\cite{Yuval:2018:Mastik};  L1 and TLB Evict+Time attacks~\cite{Gras:2018:TLB,Osvik:2006:cacheFlushing},  Rowhammer~\cite{rowhammer:2020:github,Mutlu:2014:Rowhammer},  Meltdown~\cite{meltdown:2020:github,Lipp:2018:Meltdown}. The entire framework just adds around 20 lines of C code in the Linux kernel to provide micro-architectural security for the entire system. 
    \item We have evaluated \name with multiple benchmark suites including the
    SPEC-2017~\cite{spec2017}, SPEC-2006~\cite{SPEC2006},  SPECViewperf-13~\cite{specviewperf13} and STREAM~\cite{McCalpin:2007:stream} and found  an average perfomance overhead of 1\%.

\end{itemize}

The rest of the paper is organized as follows. Section~\ref{sec:background} presents the necessary background on micro-architectural attacks, Hardware Performance Counters, and Linux scheduler. In Section~\ref{sec:relatedWork}, we discuss the related works in micro-architectural attack mitigation. Section~\ref{sec:keyIdea} sums the key idea in the~\name framework. 
In Sections~\ref{sec:implementation} and~\ref{sec:results} we discuss the implementation details and present the evaluation and results, respectively. Section~\ref{sec:caveats} discusses the caveats and limitations of~\name. Section~\ref{sec:concAndFutureWork} concludes the paper.

\section{Background} \label{sec:background}

\subsection{Micro-architectural attacks}
\label{sec:micro-architectural_attacks}

A micro-architectural side-channel attack leverages leakage from shared micro-architectural resources in a system to glean sensitive information from a process. We illustrate a micro-architectural attack that uses  a shared L1-data cache to create a covert channel between a {\em sender} and a {\em receiver} process. Apriori, the sender and receiver agree upon two cache sets for communicating 0 and 1 respectively. The receiver first performs memory operations that fill both  cache sets. This is known as the {\em prime} phase. Depending on the message bit, the sender performs a memory operation to evict the receiver's data from the corresponding cache set. In the {\em probe} phase, the receiver performs the memory operations again and times the accesses.  Based on the access time, the receiver can infer the transmitted bit since the memory access to the evicted cache set would take longer owing to the cache miss. 

Later in the paper, we use this covert channel as a running example to explain~\name and show how the transmission-rate between sender and receiver can be reduced by throttling the receiver's CPU time.
Further, we apply~\name to other micro-architectural attacks, including  Meltdown~\cite{meltdown:2020:github,Lipp:2018:Meltdown}, Rowhammer~\cite{rowhammer:2020:github,Mutlu:2014:Rowhammer}, L1 data cache attacks on AES~\cite{Osvik:2006:cacheFlushing}, L1 instruction cache attacks on RSA~\cite{Aciimez:2010:instncache}, a TLB covert channel~\cite{Gras:2018:TLB} and a prime-and-probe LLC covert channel~\cite{Yuval:2018:Mastik}.

\subsection{Hardware Performance Counters}

Most modern processors have a Performance Monitoring Unit on-chip to monitor micro-architectural events of running applications. Each logical core has a dedicated set of  4 to 8 configurable registers that can count the number of times a particular event occurs in a given duration. These registers are called Hardware Performance Counters (HPCs) and can be used to monitor a wide range of events like CPU-cycles, cache accesses, context-switches, and page faults. The number of such events that can be monitored simultaneously  is limited by the number of HPCs available in the hardware.

Patterns present in the time series traces of performance counter events have been shown  to characterize the behavior of a thread's execution. This has been leveraged in several applications such as  detecting  anomalous behavior in programs~\cite{Krishnamurthy:2020:anomaly,Tang:2014:anomalyHPC,Woo:2018:Anomaly}, side-channel attacks~\cite{Aweke:2016:anvil,Chiapetta:2016:hpcdetection,Mushtaq:2018:NightsWatch,Zhang:2016:cloudRadar} and   malware analysis~\cite{Basu:2020:hpcTheoretical,Demme:2013:feasibility,Tang:2014:anomalyHPC}. \nameA, similarly uses HPCs to compute a {threat index} for a thread. Threads which depict a micro-architectural attack like behavior are given a high {threat index} and are throttled by reducing their CPU time.

\subsection{Process Scheduling}\label{sec:background_scheduler}
~\name modifies the OS scheduling algorithm to consider the threat index of a thread for CPU resource allocation. While we can incorporate~\name in most scheduling algorithms, in this paper, we demonstrate ~\name with the Completely Fair Scheduler (CFS);  the default scheduler present in the Linux kernel since Version 2.6~\cite{linuxkernel:2019}.

The CFS algorithm tries to achieve the ideal multitasking environment where threads with equal priorities receive the same share of processor time, called {\em timeslice}. The {timeslice} allocated to a thread ${\tt t}$, denoted $\Delta_{\tt ts}$, is a fraction of a  predefined value called {\em targeted latency} ($\Delta_{\tt tl}$).   When multiple threads compete for CPU time, the scheduler  allocate timeslices in proportion to a metric called {\em weight} of the thread as follows
\begin{equation}
    \Delta_{\tt ts} = \Delta_{\tt tl} \times  \frac{\tt w}{\sum_{\tt threads} {\tt w}} = \Delta_{\tt tl} \times {\tt s} \enspace,
    \label{eqn:ts}
\end{equation}
where ${\tt w}$ is the weight of the thread ${\tt t}$, ${\sum_{\tt threads} \tt w}$  is the sum of weights of all the threads sharing the CPU, and ${\tt s}$ is the {\em relative weight} of  thread ${\tt t}$. On thread creation, its {weight}  takes a default value (${\tt w_{\tt DEF}}$) which lies in the middle of 40 discrete levels, ranging from ${\tt w_{MIN}}$ to ${\tt w_{MAX}}$. The ratio of weights at two consecutive levels $\gamma$, $(0 < \gamma < 1)$ is determined by the OS scheduler at design time. The relationship between these weight levels can be given as
${\tt w_{MIN} = {\gamma}^{19} w_{DEF} = {\gamma}^{39} w_{\tt MAX}}$.
The {weight} also gives a notion of thread priorities. A higher weight value for thread implies a larger timeslice and a higher frequency of getting scheduled for execution.

\section{Related work} \label{sec:relatedWork}
There is a pressing need to develop lightweight, portable solutions for micro-architectural attacks. Most preventive countermeasures that
have low overheads require hardware modifications~\cite{Domnister:2012:nonMonopolizableCache,intel:2015:CAT,Kong:2009:Hardware-Software,Liu:2016:CATalyst,Oberg:2013:EDAtestingSolution,Page:2005:cachePartitioning,Qureshi:2018:Ceaser,Qureshi:2019:CS,Tiwari:2009:executionLeases,Wang:2008:randomizedCaches,Yan:2018:invisiSpec} and cannot be incorporated in existing systems. To achieve portability, there have been a few attempts to tackle  micro-architectural attacks from the Operating System, such as flushing shared resources~\cite{Cock:2013:latticeScheduling,hu:92} and soft isolation of Virtual Machines sharing the same hardware~\cite{Varadarajan:2014:schdulerCrossVMSCdefense}. Nomani and Szefer in~\cite{Nomani:2015:schedHPC} incorporate a neural-network within the OS scheduler, to detect tasks requiring the same functional units and scheduling them on different cores to minimize contention. Such solutions are likely to have scalability issues as the workloads increase.

Most contemporary detection-based solutions use HPCs and Machine Learning (ML) models  ~\cite{Alam:2017,Chiapetta:2016:hpcdetection,Mushtaq:2018:NightsWatch,mushtaq:2018:prime+probe,Mushtaq:2020:whisper}. Among the  several shortcomings~\cite{Das:2019:sokHPC} of these works, 
the most serious limitation  is with the false positives which  adversely affect benign program execution. Contemporary works  use complex techniques like KNNs, decision trees, neural networks and ensemble methods to reduce false positives which can still have unacceptable impacts. Furthermore, the complex techniques used have considerable overheads and could  reduce the prediction rate leading to  fast attacks completing undetected.
Unlike these works, \name reacts to a flagged program by throttling its CPU-share, thus preventing the attack from completing. Contemporary works, on the other hand, would either migrate the program to another CPU or terminate it. The advantage we achieve with the feedback loop of \name is that falsely flagged threads can recover and regain their CPU-share with only slight increase in execution time. Further, due to the reactive control loop, very simple detection mechanisms are sufficient.

\section{The ~\name Framework} \label{sec:keyIdea}
\begin{figure}[!t]

 \center

  \includegraphics[width=\columnwidth]{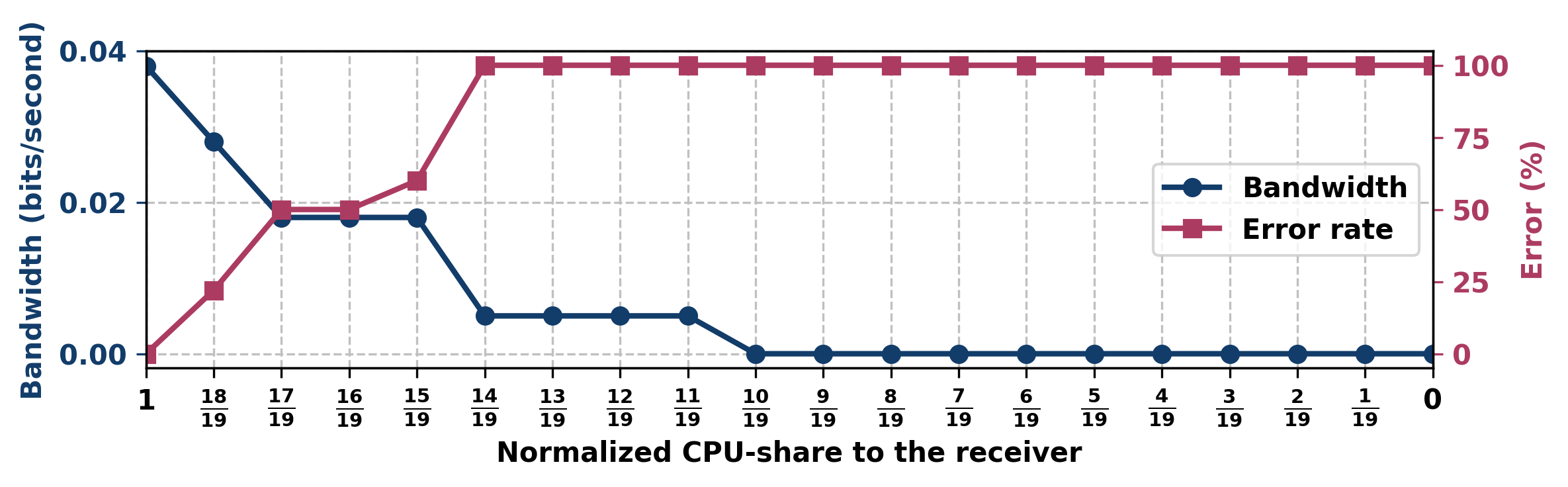}\vspace{-0.1cm}
  
  \caption{Effects of CPU-share to the {\em receiver} thread on the L1-data cache covert channel's Bandwidth and Error rate.~\name uses this as a premise to throttle attack programs on detection.}
\label{fig:covert_channel_throttle}
\vspace{-0.2cm}
\end{figure}

\begin{figure}[!t]{
  \includegraphics[width=\columnwidth]{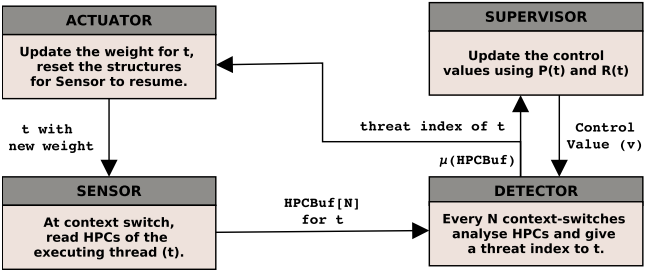}  }
  \vspace{-0.4cm}
\caption{The~\name reactive loop. The Sensor logs the HPC values for a thread ${\tt t}$ in ${\tt HPCBuf}$ 
and passes it to the Detector which computes the  threat index for ${\tt t}$. It then passes the  threat index and the mean of ${\tt HPCBuf}$ to the Supervisor which adjusts the control value $({\tt v})$ for the next epoch, and to the Actuator which updates the weight of ${\tt t}$ as per its threat index. This  loop continues throughout the runtime of ${\tt t}$.}
\label{fig:leash_cycle}
  \vspace{-0.1cm}
\end{figure}

Micro-architectural attacks depend considerably on CPU resources. If the attack programs are starved of the CPU, the success drops considerably. Consider, for example, the L1-cache timing covert channel discussed in Section~\ref{sec:micro-architectural_attacks}. Figure~\ref{fig:covert_channel_throttle} shows that starving the receiver thread not only reduces the communication bandwidth considerably but also increases the transmission errors. The reduced bandwidth and the increased error is due to smaller CPU-time shares, which prevents the receiver from  performing a sufficient number of timing measurements.

Along with being highly CPU-bound, micro-architectural attacks are procedural and repetitive by design, which gives them a  distinct execution behavior. For example, the receiver in the cache covert channel continuously executes a small set of instructions in a loop performing extensive memory operations. Due to the continuous probing, the number of memory accesses by the receiver increases to be significantly higher compared to a regular thread. ~\name uses  HPCs to detect such anomalous behavior and penalizes such threads by decreasing their weight, which in turn reduces their timeslice (Equation~\ref{eqn:ts}). If the thread stops exhibiting the anomalous behavior, its weight is gradually increased, thus regaining its regular timeslice. To achieve this,~\name has four components: a Sensor, {Detector $\mathcal{D}()$}, {Actuator $\mathcal{A}()$}, and a Supervisor $\mathcal{S}()$ as shown in Figure~\ref{fig:leash_cycle}. 

\vspace{-0.2cm}

\begin{figure}[!t]{
  \includegraphics[width=\columnwidth]{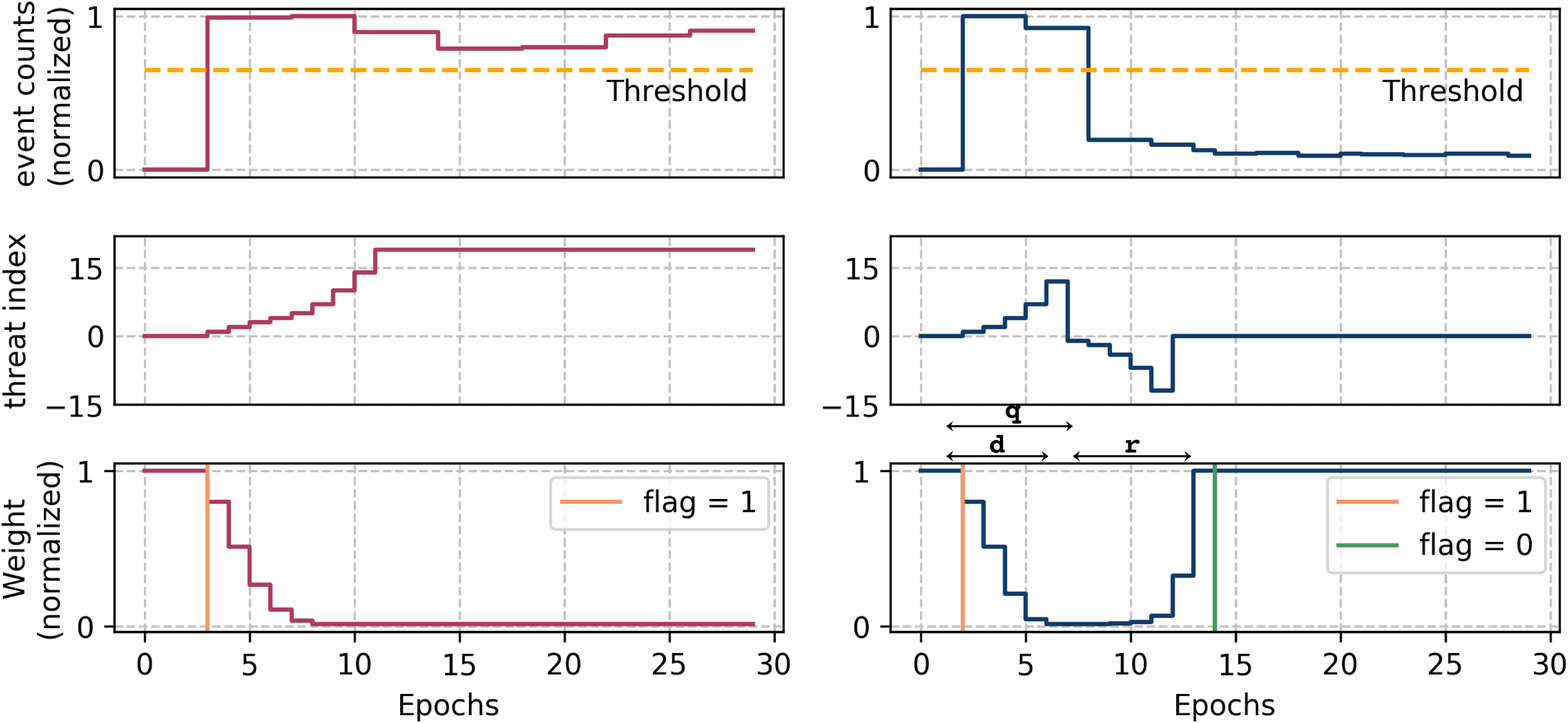} }
  \vspace{-0.4cm}
\caption{~\name framework effects on an L1-data cache covert channel  (left column) and a benign program bzip2 (right column) from the SPEC-2006 benchmark suite.} \vspace{-0.2cm}
\label{fig:covert_vs_bzip2}
\end{figure}

{\flushleft \bf Sensor.} 
At every context switch,~\name reads the values of HPCs and logs the event counts since the last context switch. The log is stored in a buffer of size $\tt N$, called ${\tt HPCBuf}$,  in the context of the scheduled-out thread. Thus, every thread executing in the CPU has a time-series trace for each HPC  logged during its execution.  For example, the first row  in Figure~\ref{fig:covert_vs_bzip2} shows the partial time-series traces for the L1-D covert channel described in Section~\ref{sec:micro-architectural_attacks} and the {\tt bzip2} application from the SPEC-2006 benchmark suite. The counter corresponds to an event that counts the delay cycles due to DSB (Decoded Stream Buffer) to MITE (Micro Instruction Translation Engine) switches on an Intel i7-3770 processor. 

The counters to log are determined prior to the deployment of ~\name based on the execution characteristics of the attack vectors considered. The number of counters that can be logged is bounded by the number of available HPC registers in the hardware. Reducing the number of HPCs logged would reduce context-switch overheads, while increasing the number of HPCs logged would widen the scope of the attacks that can be detected. Section~\ref{sec:implementation} discusses this further.

\vspace{-0.2cm}

\begin{algorithm}[!t]
\DontPrintSemicolon
{\bf Initial State:} At ${\tt i = 0}$, ${\tt T_{i}} = 0$ and $\tt flag = 0$.
{\bf Global: }Event count threshold: $\tau$, default weight: ${\tt w_{DEF}}$.
\KwIn{
Thread ${\tt t} = \{ {\tt w_{i-1}} ,  {\tt T_{i-1}}, {\tt HPCBuf} \}$.}
\KwResult{Threat index for $i$-th epoch: ${\tt T_i}$, ${\tt flag}$}
\Begin{
$\mu = \text{mean}({\tt HPCBuf})$~\\
\If{${\tt flag} = 1$}{
${\tt T_i}$ = ${\tt T_{i-1}}$ + $\mathcal{S}({\tt t})$ \tcp{\tt  threat index update}~\\
\If{${\tt w_{i-1}} = {\tt w_{DEF}}$}{ ${\tt flag} = 0$ \tcp{\tt Thread ${\tt t}$ is unflagged}\label{cmt:unflagged}
    ${\tt T_i = 0}$ }} 
\If{(${\tt flag=0}$ {\bf and} $\mu \ge \tau$ ) }{
    {${\tt flag} = 1$} \tcp{\tt Thread ${\tt t}$ is flagged}\label{cmt:flagged}}
\KwRet{ $({\tt T_i, flag})$}}
\caption{Detector component in~\name for thread ${\tt t}$ at the beginning of the $i$-th epoch.}\label{algo:detector}
\end{algorithm}

{\flushleft \bf Detector.} 
At every ${\tt N}$ context switches  (called an {\em epoch}) for the thread, the Detector  $\mathcal{D}()$ is triggered and uses Algorithm~\ref{algo:detector}  to compute the new threat index  for the thread.  The threat index in the $i$-th epoch $(\tt T_i)$ depends on the mean of the ${\tt HPCBuf}$ values ($\mu$), the event-specific  threshold ($\tau$), the threat index of the previous epoch ${\tt T_{i-1}}$ and the control value ${v_i}$ passed by the Supervisor function $\mathcal{S}()$ as,
\vspace{-0.2cm}
\begin{equation}\label{eqn:detector}
    {\tt T_i} = \mathcal{D}({\tt { HPCBuf, T_{i-1}}, v_{i}}) \enspace .\vspace{-0.2cm}
\end{equation}

\name flags threads which begin to behave maliciously (Line 11 in Algorithm~\ref{algo:detector}). This happens when the mean $\mu$ of an unflagged thread first violates the threshold. For every subsequent epoch for this thread, where $\mu$ exceeds the threshold, its threat index is modified as per the output of the Supervisor function $\mathcal{S}()$ described in Equation~\ref{eqn:supervisor_step_function} (Line 6). \name allows a flagged thread to recover if it stops behaving maliciously. The recovery starts when $\mu$ for the flagged thread first falls below the threshold. The thread's threat index is then  reduced in subsequent epochs until the thread reattains the default weight, which restores its CPU share. When this happens, the thread is unflagged and its threat index is reset (Lines 8 and 9).

Figure~\ref{fig:covert_vs_bzip2} shows two threads, a L1 Data Cache Covert Channel and a benign program {\tt bzip2} from the SPEC-2006~\cite{SPEC2006} benchmark suite. While both  threads, with counter values greater than the threshold get 
flagged by \name, the benign thread {\tt bzip2} recovers its fair share of CPU resources (in the 14th epoch) when its counter values go below the threshold. The CPU share for the covert channel is never restored as its counter values always remains over the threshold. Thus the covert channel is never unflagged. Its threat index saturates when the minimum weight is reached.

\vspace{-0.2cm}
{\flushleft \bf Actuator.} For each thread in the $i$-th epoch, the Actuator receives the threat index $({\tt T_i})$ from the Detector and modifies the thread's weight ${\tt w_i}$ and in turn, the relative weight ${\tt s_i}$ (Equation \ref{eqn:ts}) for the thread. The relative weight for the thread at the beginning of the $i$-the epoch is given as
\vspace{-0.1cm}
\begin{equation}\label{eqn:actuator}
    {\tt s_i} = \mathcal{A}({\tt s_{i-1}, T_{i}}) = {\tt s_{i-1} - \gamma ({\tt s_{i-1}})\times T_{i}}\enspace, \vspace{-0.2cm}
\end{equation}
where is ${\gamma}$, (0 < $\gamma$ < 1)  is a constant fixed by the OS scheduler which  determines the amount of fall in the weight with every increase in the threat index. In our evaluation platform, $\gamma = 0.1$ which means that, for every rise in threat index values, the weight drops by 10\% until it reaches ${\tt w_{MIN}}$. Similarly, when a thread is recovering, the threat index value is negative and hence every fall in threat index value increases its weight by 10\% until its weight is restored. 
The adaptable design of~\name  efficiently brings down the cost of a false penalization. Once a benign thread, which is erroneously flagged, is unflagged, it regains its CPU share and executes without any additional overheads.

\vspace{-0.2cm}

{\flushleft \bf Supervisor.} 
\nameA's Detector identifies threads that are penalized and rewarded. Ideally, this decision should be based on careful analysis of the ${\tt HPCBuf}$ data so that malicious threads are quickly identified and penalized while benign threads are not affected. However, supporting sophisticated analysis in the Detector is difficult because it executes in a context switch and therefore affects the performance of the entire system. Thus, as seen in Algorithm~\ref{algo:detector}, the Detector is kept very simple and comprises of about 6 instructions  
that are executed at every epoch (${\tt N}$ context switches).

To support complex analysis on the performance counter data, \name uses a Supervisor that does not need to execute in the context switch. The Supervisor provides thread specific control value $({\tt v_i})$ that represents the amount by which the thread index ${\tt T_i}$ of a flagged thread is degraded or upgraded in the $i$-th epoch. The output of the $\mathcal{S}({\tt t})$ for a thread ${\tt t}$ is given by
\vspace{-0.1cm}
\begin{equation}\label{eqn:supervisor}
    {\tt v_i} = \mathcal{S}({\tt v_{i-1}, T_{i-1}, T_{i-2}, T_{i-3}, \cdots} )\enspace.
\end{equation}

A general framework for the Supervisor function for thread ${\tt t}$ involves a penalty function  ${\tt P(t)}$  and reward function ${\tt R(t)}$, as given in Equation(~\ref{eqn:supervisor_step_function}).  Such a design allows the Supervisor to be implemented with different policies
\vspace{-0.2cm}
\begin{align}
    \mathcal{S}({\tt t}) = \begin{cases}
    {\tt P(t)}, & \text{$\mu$ > $\tau$ (${\tt t}$ is malicious)}\\
    {\tt R(t)}, & \text{${\tt t}$ is flagged and $\mu$ $\leq$ $\tau$ (recovering)}\\
    0, & \text{thread ${\tt t}$ is unflagged, }\label{eqn:supervisor_step_function}
    \end{cases}
\end{align}
where $\mu$ is the mean of ${\tt HPCBuf}$ and $\tau$ is the threshold. Configurable Supervisor policies allow~\name to handle new attack vectors. Section~\ref{sec:spervisorPolicy} provides more details about the Supervisor. 

\section{Implementation} \label{sec:implementation}
This section provides the implementation details of \name and some of its primary design choices.

\subsection{Sensing Program Behavior}

Most modern processors have 4 to 8 {\em Hardware Performance Counters} (HPCs) per logical core that can be configured to count one of several micro-architectural events such as cache misses, branch events, or loads and stores to various levels of memory. To enhance readability, we label 40 selected events as ${\tt e_1, e_2, \cdots, e_{40}}$ in the paper. We describe these events and associated masks defined by the micro-architecture in detail in the Appendix.

\subsection{Enhancing the CFS Scheduler for \name}

\begin{algorithm}[!t]
\DontPrintSemicolon

\KwIn{Task ${\tt prev}$ getting preempted from runqueue ${\tt rq}$ by task ${\tt next}$, per thread counter ${\tt cs\_count}$ to log the number of context switches.}
\KwResult{${\tt rq}$: ${\tt prev}$ reweighed, ${\tt next}$ scheduled,  Logged HPC data for both}
\Begin{
\If{ \text{threads in} ${\tt rq = 1}$ {\bf and } ${\tt flag=1}$ } {
wake up dummy thread}
\If{ ${\tt prev.start}$ } 
{
${\tt prev.end}$  $\leftarrow$ {read event count}
${\tt prev.HPCBuf[prev.cs\_count]}$~\\ \quad \quad \quad  ${\tt \leftarrow (prev.end - prev.start)}$~\\
increment ${\tt prev.cs\_count}$ }
\If{${\tt prev.cs\_count == N}$} {
${\tt Detector(prev)}$ \quad \tcp{Algorithm~\ref{algo:detector}}\label{cmt}
Reset ${\tt HPCBuf}$, ${\tt prev.cs\_count}$
}
{CFS context switching steps }\label{cmt}~\\
${\tt next.start \leftarrow}$ {read event count}~\\
\KwRet{ ${\tt rq}$} }
\caption{Context Switching in \name}\label{algo:context_switch}
\end{algorithm}

In the Linux kernel's CFS implementation, the context switch is implemented in a function called ${\tt context\_switch}$\footnote{This function is defined in {\tt kernel/sched/core.c} in the kernel source.}. The function switches context from a thread called ${\tt prev}$ to a thread called ${\tt next}$ as shown in Algorithm~\ref{algo:context_switch}. To support~\nameA, the ${\tt context\_switch}$ function is modified to perform three additional operations;
{\em (a)} it schedules a {\em dummy} thread if needed;
{\em (b)} it logs the performance counter data in the scheduled out thread's task structure; and
{\em (c)} it executes the detector when the ${\tt HPCBuf}$ gets full. We describe each of these operations.

{\flushleft \bf Dummy Thread.} As described in Equation (\ref{eqn:ts}), the targeted latency $\Delta_{\tt tl}$ is shared among the threads in the Run Queue ($\tt rq$), proportional to their weights. However, when there is only one thread in the Run Queue, the timeslice it receives is equal to $\Delta_{\tt tl}$ and is independent of the weight. Penalizing the thread, in this case, does not affect the timeslice.

To solve this,~\name uses a special thread called {\em dummy thread}. The thread starts when the system boots, runs an infinite loop in privileged mode. It is  signaled to wake up only when the Run Queue has exactly one thread and that is flagged (Lines 2-3 in Algorithm~\ref{algo:context_switch}). In all other cases, the dummy task continues to sleep. The dummy task thus ensures that the weights of a thread always influence its timeslice.

\begin{figure}[!t]

 \center

  \includegraphics[width=\columnwidth]{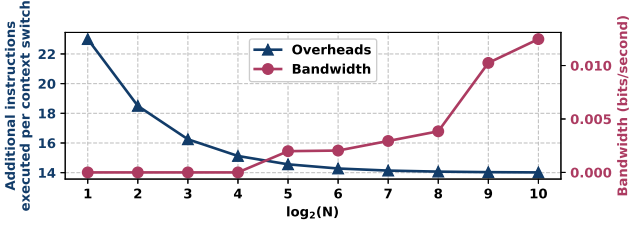}

  \caption{The number of additional instructions executed per context switch and L1-data cache covert channel bandwidth with different values ${\tt N}$. }
\label{fig:N_and_overhead}
\end{figure}

\begin{table*}[!t]
\small
\begin{center}
\caption{ Micro-architectural attacks evaluated; their leaking component; and the attack strategy used. For our evaluation we have used ${\tt e_2}$, ${\tt e_{11}}$, ${\tt e_{12}}$, and ${\tt e_{39}}$.\label{tab:attacklist}}
\begin{tabular}{|M{.22\textwidth}|M{.083\textwidth}|M{.11\textwidth}|M{.51\textwidth}| } 
 \hline
 {\bf Attack} & {\bf Leaking} & {\bf Strategy} & {\bf Example Event(s)}  \\
    & {\bf Component} & & {\bf to detect the attack} \\ 
  \hline
 L1-D cache Covert Channel~\cite{Osvik:2006:cacheFlushing} &  L1-D cache & Prime+Probe & ${\tt e_{12}}$: cycles for which the instruction decode queue is empty, or ${\tt e_{26}}$: writebacks from L1D to L2 cache lines in modified coherency state. \\ 
 \hline
  L1-D cache attack on AES~\cite{Osvik:2006:cacheFlushing} &   L1-D cache  & Evict+Time & ${\tt e_{12}}$, or ${\tt e_{28}}$: L2 store read for ownership (RFO) requests from the thread.\\ 
 \hline
 L1-I cache attack on RSA \cite{aciicmez:2007:icache,Aciimez:2010:instncache} &  L1-I cache & Prime+Probe &  ${\tt e_{39}}$: the number of prefetcher requests that miss the L2 cache, or ${\tt e_{13}}$: the number of instruction cache misses,\\ \hline
 LLC Covert Channel \cite{Yuval:2018:Mastik}  &  LLC  & Prime+Probe & ${\tt e_{39}}$, or ${\tt e_{37}}$: the cache misses for references to the LLC \\ 
 \hline
   Meltdown \cite{meltdown:2020:github,Lipp:2018:Meltdown} & Memory & Flush+Reload & ${\tt e_2}$: the mispredicted branch instructions at retirement, or ${\tt e_{15}}$:  direct and indirect near call instructions\\ 
 \hline
 Rowhammer \cite{rowhammer:2020:github,Mutlu:2014:Rowhammer} & DRAM & Hammer access & ${\tt e_{2}}$, or ${\tt e_{17}}$: the dirty L2 cache lines evicted by demand.\\ 
 \hline
   TLB Covert channel \cite{Gras:2018:TLB} & TLB & Evict+Time & ${\tt e_{11}}$: the misses in all TLB levels\\ 
 \hline
\end{tabular}

\end{center}
\end{table*}
{\flushleft \bf Log Buffer.} The log buffer, ${\tt HPCBuf}$, is a per-thread buffer of size ${\tt N}$.  It logs the event counter values and when full, marks the end of an epoch and triggers the Detector. The value of ${\tt N}$ directly affects the overheads and effectiveness of~\nameA. Figure~\ref{fig:N_and_overhead} shows how the average overhead per context switch decreases with an increase in ${\tt N}$. This is because a small ${\tt N}$, implies more frequent calls to the Detector (Lines {9-10} in Algorithm~\ref{algo:context_switch}), increasing the amortized number of additional instructions executed per context-switch. 

A small value of ${\tt N}$ also allows~\name to perform the detection of malicious threads at a finer granularity. Figure~\ref{fig:N_and_overhead} shows that as ${\tt N}$ increases, the bandwidth of the covert channel also increases. This is because the coarse detection granularity allows the covert channel to execute for sufficient time to transfer information from sender to receiver. As seen in the Figure~\ref{fig:N_and_overhead}, a buffer size between 16 and 64 entries provides a good trade-off between the overheads and the effectiveness of~\nameA.

\begin{figure}[!t]{
  \includegraphics[width=\columnwidth]{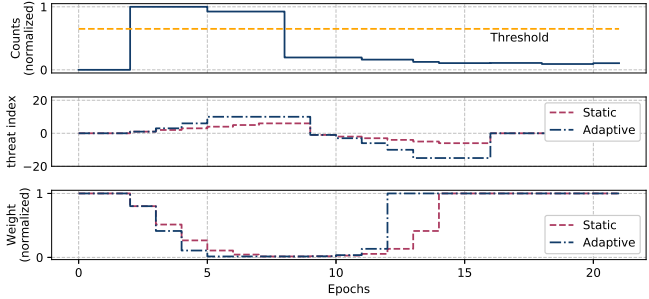}}

\caption{The threat index and weight of ${\tt bzip2}$ violating the event threshold for a few epochs, with Static and Adaptive Supervisor policies. 
}
\label{fig:bzip2_static_vs_dynamic} 
\vspace{-0.2cm}

\end{figure}

\subsection{Supervisor Policy\label{sec:spervisorPolicy}}
The Supervisor  in~\name manages the control values, as shown in Equation (\ref{eqn:supervisor_step_function}).  Algorithm~\ref{algo:detector} shows how these  values affect the threat index and, in turn, the weight of a flagged thread. The key components of the Supervisor are the penalty (${\tt P(t)}$) and  reward (${\tt R(t)}$) functions that respectively provide the rate of degradation and recovery for a flagged thread.

The penalty and reward functions are pluggable. The simplest form is a {\em static policy}, where the outputs of the functions are constants. Such a policy is agnostic to the  thread execution history.
In contrast, an {\em adaptive policy} uses the thread execution history. An  example of such policy is ${\tt P(t)= P(t)+ 1}$ and ${\tt R(t)= R(t) + 1}$ which increments the penalty values every flagged epoch and the reward value every recovering epoch.  Figure~\ref{fig:bzip2_static_vs_dynamic}, shows the difference between the two policies for a fragment of the benign {\tt bzip2} execution, where the counter briefly crossed the threshold for seven epochs. With the static policy, the { threat index} is increased with the same rate, independent of the previous epochs. Similarly, when the counter drops below the threshold, the { threat index} decreases with the same rate until the thread is unflagged. With the adaptive policy, the output of $\tt P(t)$ and hence the rate of increase of threat index (Algorithm~\ref{algo:detector}), is incremented in every flagged epoch. Similarly, when the counter drops below the threshold, the {threat\_index} is reduced much more quickly.  As seen in Figure~\ref{fig:bzip2_static_vs_dynamic}, recovery for this benign thread takes 2 epochs less compared to the static policy.

\section{Evaluation and Results} \label{sec:results}

\vspace{-0.2cm}

Every micro-architectural attack is characterized by two aspects. The first is the micro-architectural component that leaks information, for example, shared resources like the L1 data~\cite{Bernstein:2005:CacheAttack,Osvik:2006:cacheFlushing} and instruction~\cite{Aciimez:2010:instncache} caches, the Last Level Cache (LLC)~\cite{Liu:2015:LLC,Yarom:2014:flushReloadLLC,Yuval:2018:Mastik},  Translation Lookaside Buffer (TLB)~\cite{Gras:2018:TLB}, DRAM~\cite{Mutlu:2014:Rowhammer}, and Branch Prediction Units (BPU)~\cite{Evtyushkin:2018:Branchscope, Sarani:2015:watchmen}. The second is the attack strategy used to extract information from the leaking resources. Several strategies such as Prime+Probe~\cite{aciicmez:2007:icache,Aciimez:2010:instncache,Liu:2015:LLC}, Evict+Time~\cite{Himanshi:2019:SpyCartel,Hund:2013:againstASLR},  Flush+Reload~\cite{Yarom:2014:flushReloadLLC}, and time-driven attacks~\cite{ Bernstein:2005:CacheAttack,Evtyushkin:2018:Branchscope} have been proposed for this purpose. We evaluate \name with seven different micro-architectural attacks chosen to get a wide coverage of leaking components and attack strategies. Table~\ref{tab:attacklist} provides the details of attacks evaluated with some example events to detect them. Further in this section we discuss the deployment of \name on a system and present the evaluation results for various attacks. We then present the overheads on various workloads including the SPEC-2006~\cite{SPEC2006}, SPEC-2017~\cite{spec2017}, SPECViewperf-13~\cite{specviewperf13} and STREAM~\cite{McCalpin:2007:stream} benchmark suites.
\begin{figure*}[!t]

 \center

  \includegraphics[width=\linewidth]{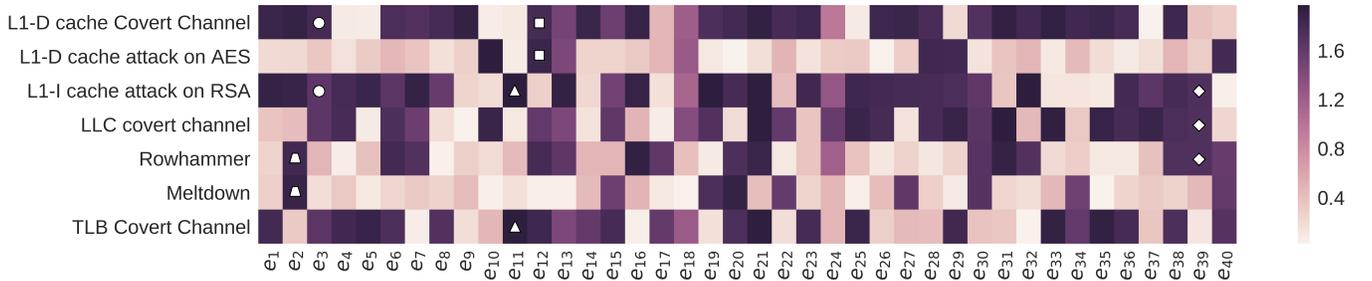}
   
  \caption{A heatmap of the Detectability Score of 40 selected events for each of the attacks in Table~\ref{tab:attacklist}. Identically annotated events for multiple attacks can be used to detect all those attacks, such as event ${\tt e_2}$ for Rowhammer and Meltdown. }
  
  \label{fig:heatmap}

\end{figure*}

\begin{figure*}[!h]
\captionsetup[subfloat]{farskip=0.5pt,captionskip=0.5pt}
\centering
\begin{tabular}{ccc}

\subfloat[L1-data cache attack on AES]{\includegraphics[width=0.3\textwidth]{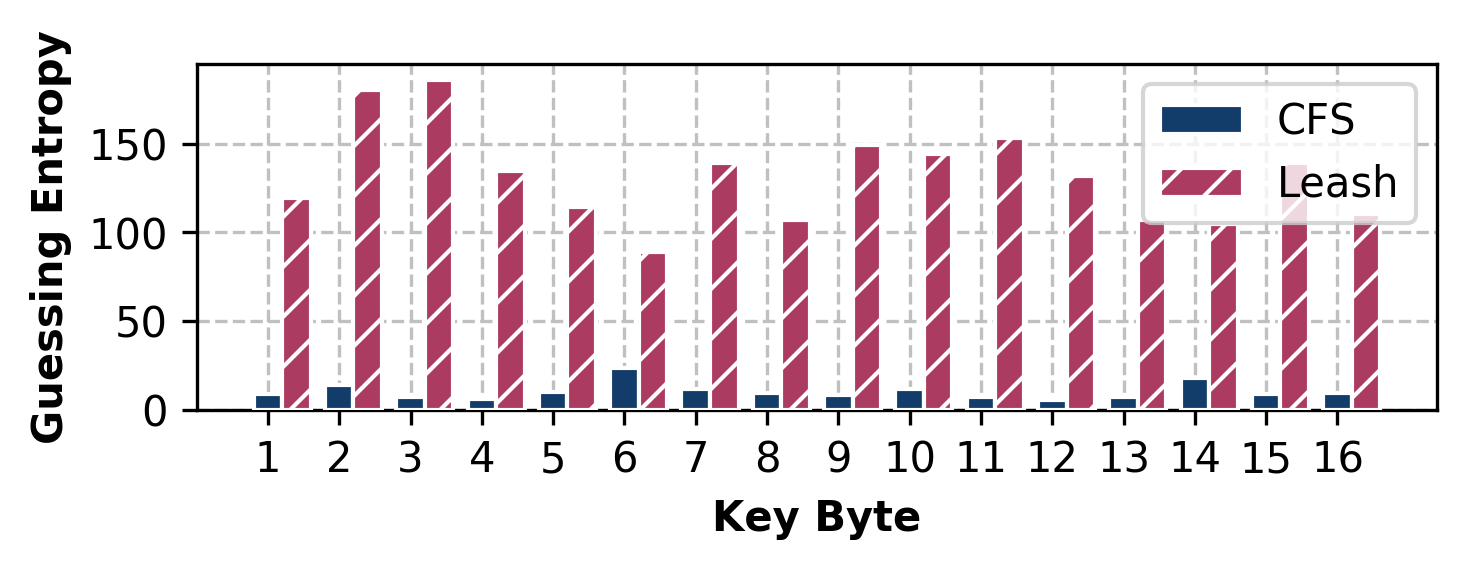}\label{fig:resAES}} &
\subfloat[L1-instruction cache attack on RSA]{\includegraphics[width=0.3\textwidth]{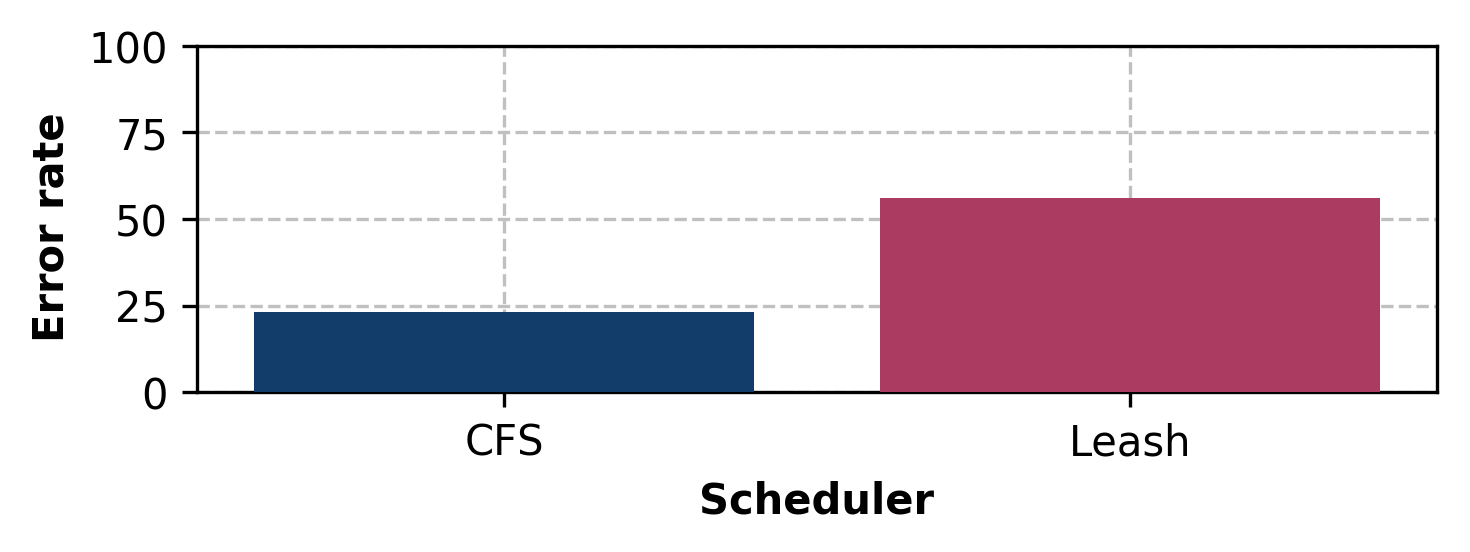}\label{fig:resL1instn}} &
\subfloat[LLC covert channel transmitting 10 bits.]{\includegraphics[width=0.3\textwidth]{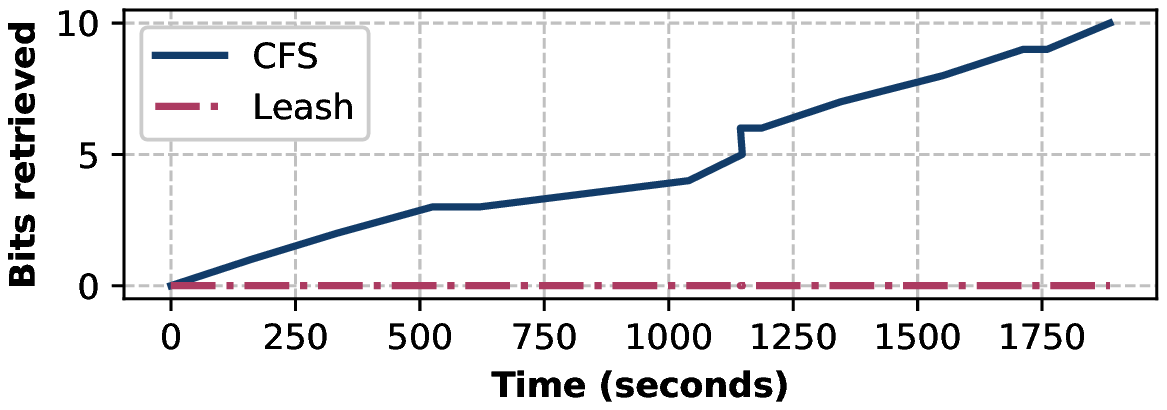}\label{fig:resLLC}} \\

\vspace{-0.2cm}

\subfloat[Rowhammer for 200 iterations]{\includegraphics[width=0.3\textwidth]{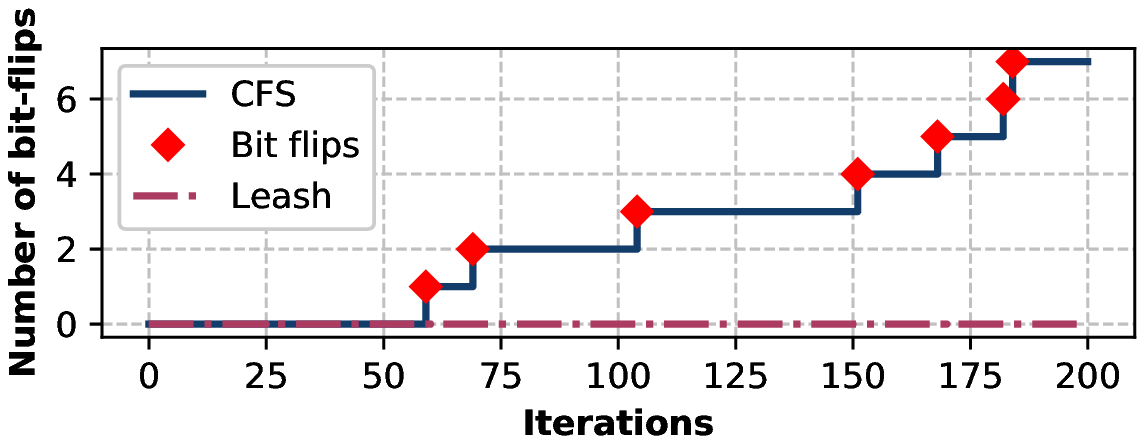}\label{fig:resrowhammer}} &
\subfloat[Meltdown]{\includegraphics[width=0.3\textwidth]{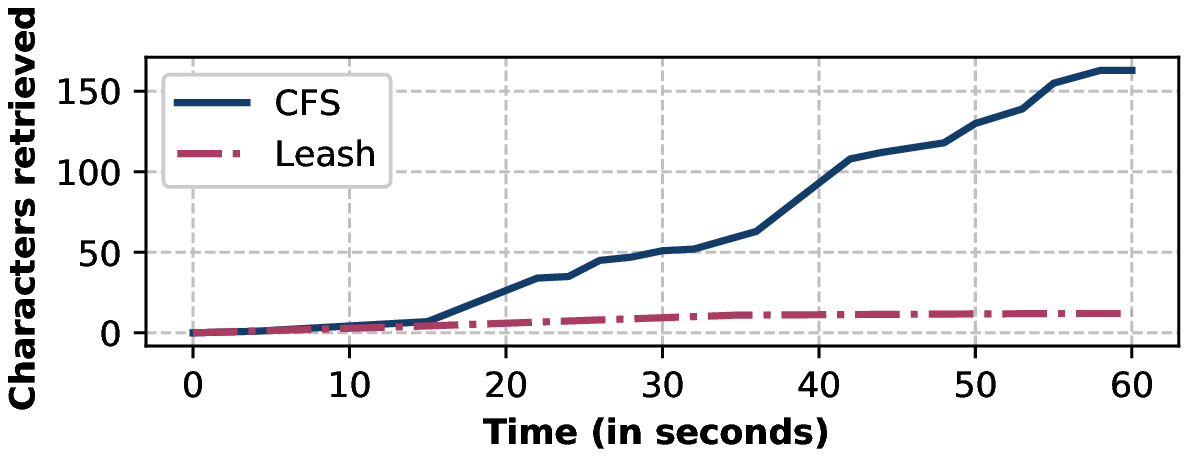}\label{fig:resmeltdown}} &
\subfloat[TLB covert channel]{\includegraphics[width=0.3\textwidth]{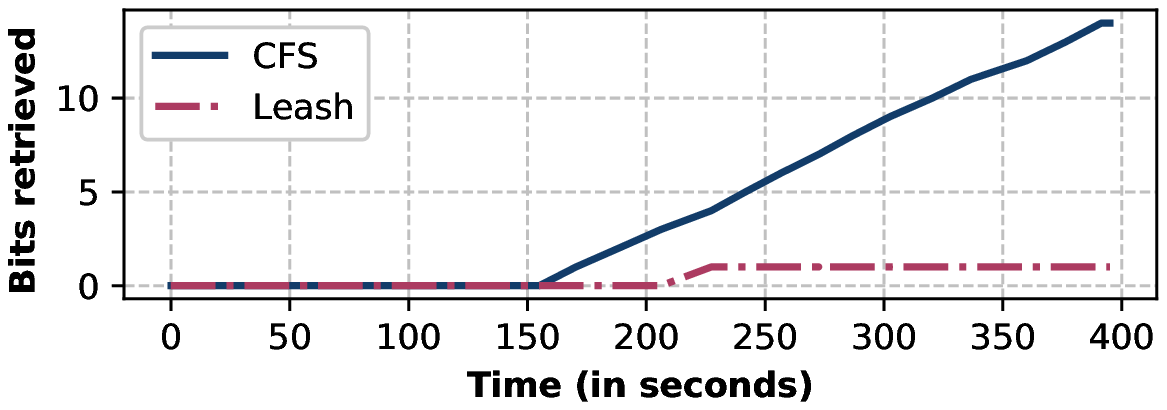}\label{fig:resTLB}} \\

\end{tabular}

\caption{The effects of~\name on different attacks.}\label{fig:attack_eval}
\end{figure*}

The evaluation is done on an Intel Core i7-3770 processor, which deploys Intel's Ivy Bridge micro-architecture. The processor has four physical cores, each with two hyperthreads. We patch the Linux kernel, version 4.19.2, with ~\name and boot Ubuntu 16.04 Operating System. We configure \name with an adaptive Supervisor policy with ${\tt P(t)=P(t) + 1}$ and ${\tt R(t)=R(t) + 1}$ (Section ~\ref{sec:spervisorPolicy}) for the evaluation.

\subsection{Deploying \name}\label{sec:multipleEvents}

Prior to the deployment of \name on a system, the designer models the  threat in terms of a set of attack vectors. Events need to be chosen to detect all attack vectors in the set. While typical microprocessors have a large number of countable events, they have a limited number of configurable Hardware Performance Counter (HPC) registers. Our evaluation platform for instance, has 208  events and four configurable HPC registers~\cite{intel:2008:intelMnaual3B}. The designer thus has to choose at most four events that can best distinguish the set of attacks from benign programs. In the subsequent sections, we present techniques to choose such events and  the evaluation results.

{\flushleft \bf Selecting HPCs.} As a representative for benign programs, we use all programs of a CPU  benchmark suite that we assume cover a variety of benign program behavior.
For each micro-architectural attack, we analyze all supported events and rank them based on distinguishability from the benign set using Principal Component Analysis~\cite{Pearson:1901:PCA}.  Figure~\ref{fig:heatmap} summarises the distinguishability score for the 7 attack vectors in Table~\ref{tab:attacklist} using SPEC-2006 CPU benchmark suite as the set of benign programs~\cite{SPEC2006} for the best 40 events of the 208 supported on our platform. Since there are only 4 HPC registers available, we configure at most 4 events that can best distinguish the 7 attacks. From Figure~\ref{fig:heatmap}, we notice that events ${\tt e_2}$, ${\tt e_{11}}$, ${\tt e_{12}}$, and ${\tt e_{39}}$ satisfy this requirement.  
Event ${\tt e_2}$ covers attacks Meltdown and Rowhammer
 event ${\tt e_{11}}$ covers attacks L1-I Cache attack and TLB Covert Channel, event ${\tt e_{12}}$ covers attacks L1-D Cover Channel and L1-D attack on AES while the event ${\tt e_{39}}$ can cover the attacks LLC Covert Channel and Rowhammer. Many other such event combinations are also possible.

{\flushleft \bf Detecting Attacks.}  \name is able to throttle all the attacks in Table~\ref{tab:attacklist}. Figure~\ref{fig:attack_eval} describes the impact of \name on these attacks.
For example, the guessing entropy~\cite{massey:94:guessing} for 1 byte of an T-table based AES key~\cite{openssl:2020:aes}, using the L1-D Evict+Time attack increases from 10 to 131 (Figure~\ref{fig:resAES}). The error in guessing 1-bit of an RSA key used in a square-and-multiply implementation~\cite{Menezes:1996:handbookCrypto} increases to 50\% (Figure~\ref{fig:resL1instn}). In both cases, the attacks are as good as an attacker that randomly guesses the key. 
The covert channels at LLC~\cite{Yuval:2018:Mastik} and TLB~\cite{Gras:2018:TLB} see a drastic fall in the number of bits communicated after getting throttled by \name (Figure~\ref{fig:resLLC} and ~\ref{fig:resTLB}). The Rowhammer~\cite{rowhammer:2020:github} attack, which induces a bit-flip in a DRAM\footnote{ Transcend  ${\tt DDR3}$-1333 645927-0350 DRAM chip.} row every 29 iterations (on average) is unable to cause a single bit-flip even after a day of execution (Figure~\ref{fig:resrowhammer}). Further, the Meltdown~\cite{meltdown:2020:github} attack that dumps  contents of the kernel memory, is throttled by \name to become ineffective (Figure~\ref{fig:resmeltdown}).

\begin{figure}[!t]
 \center
  \includegraphics[width=\columnwidth]{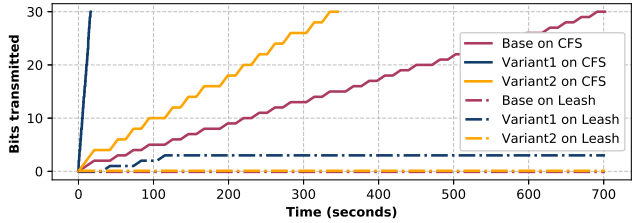}
  \vspace{-0.5cm}
  \caption{Effects of~\name on multiple variants of the L1-data cache covert channel transmitting 30 bits.}\vspace{-0.2cm}
  \label{fig:variants_l1d}
\end{figure}

{\flushleft \bf \name with   Attack Variants.} While configuring \nameA, events are chosen by profiling a specific realization of an attack. However, with the same configuration \name can detect different variants of the attacks as well. For example, the covert channel described in Section {\ref{sec:micro-architectural_attacks}}, can have multiple variants by tweaking the communication protocol. For example, 
we design a ${\tt Variant~1}$, which transmits redundant bits for reliability while a ${\tt Variant~2}$, which uses multiple sets to transmit two bits at a time. \name can detect these variants of the covert channel as seen in  Figure~\ref{fig:variants_l1d}. Without \nameA, the bandwidth is 0.43 bits/second for the base variant, 1.76 bits/second for ${\tt Variant~1}$ and to 0.087 bits/second for ${\tt Variant~2}$. With \nameA, all three covert channels are stymied and almost no bit gets transmitted correctly.

\subsection{Performance Overheads}
\begin{figure}[!t]
 \center

  \includegraphics[width=\linewidth]{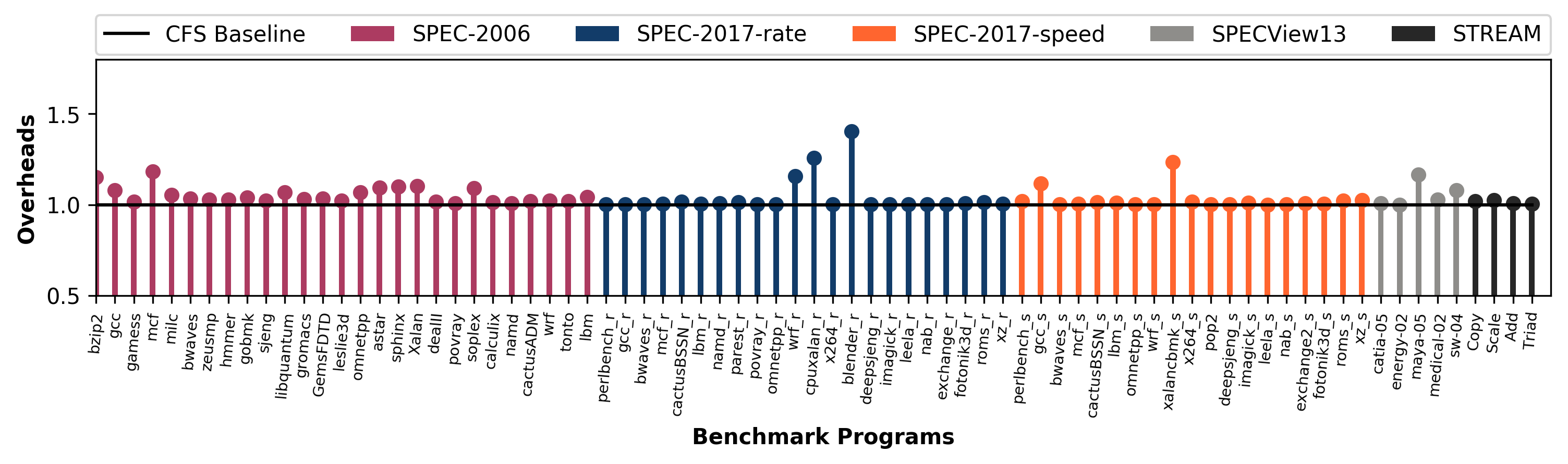}
  \caption{Performance overheads due to \name on various benchmark suites.}
  \vspace{-0.2cm}
  \label{fig:perfOverhead}

\end{figure}

We evaluated \name with several benchmark suites including SPEC-2006~\cite{SPEC2006}, SPEC-2017~\cite{spec2017}, SPECViewperf-13~\cite{specviewperf13} and STREAM~\cite{McCalpin:2007:stream}. 
SPEC-2006 and SPEC-2017 are CPU benchmark suites with different integer and floating-point programs like Machine Learning algorithms. SPECViewperf-13 is a collection of graphics-oriented benchmarks programs, while STREAM is designed to perform memory-intestive tasks.
The overheads seen in Figure~\ref{fig:perfOverhead} is due to {\bf (1)} the few additional  instructions in each context switch (Figure~\ref{fig:N_and_overhead}) and {\bf (2)} benign threads being falsely flagged. In contrast to contemporary detection counter-measures~\cite{Alam:2017, Aweke:2016:anvil,Briongos:2018:cacheShield,Chiapetta:2016:hpcdetection,Mushtaq:2018:NightsWatch,Zhang:2016:cloudRadar}, all falsely flagged programs recover and are not adversely affected.
The only case where over 25\% overheads occurs is {\tt blender\_r}, a 3D rendering program, which is  falsely flagged in 30\% of its epochs due to the prolonged malicious looking behavior. 
Out of the programs evaluated, over 45\% were flagged at-least once. However, run-time overheads on average across all  benchmarks remain low at 1\%. 
Another reason for the low overheads is that \name executes during a context switch in the kernel. Unlike
~\cite{Aweke:2016:anvil,Briongos:2018:cacheShield,Chiapetta:2016:hpcdetection,Mushtaq:2018:NightsWatch,Zhang:2016:cloudRadar}, it does not require a background task that needs scheduling and is  independent of the system load.

\section{Caveats and Limitations} 
\label{sec:caveats}
\begin{itemize}[wide,labelwidth=!,labelindent=0pt]

    \item {\em Time-Driven Attacks.} 
    Unlike the attacks discussed in Section~\ref{sec:results}, in a time-driven attack, the attacker uses the victim's execution time~\cite{Sarani:2015:watchmen,Chester:2012:boostingCacheTiming,Chester:2015:time-driven}.\name can not detect such attacks because of the absence of a spy thread to flag.  
  \item {\em High-resolution Attacks.} If a high-resolution attack like~\cite{Irazoqui:2014:flush+flush} works in less time than an epoch, \name would not be able to stop it  unless the epoch is shortened. However, this can increase the overheads (Figure~\ref{fig:N_and_overhead}). 
  \item {\em Undetectable Attacks.} While the attacks evaluated were identifiable by HPCs, there may be attacks that evade them. \name would not be able to thwart such attacks.
  \item {\em Complex adversarial attacks on \nameA.} While the Supervisor component in \name can be designed to detect multiple adversarial attacks, there may be complex strategies such as using distributed colluding threads ~\cite{Gullasch:2011:cachegames}. The Supervisor would need to be enhanced to handle such distributed attacks.
  \item {\em Detecting a large number of attacks.} As the attack vectors to detect increase, the number of events to be counted may exceed the HPCs available in the hardware, requiring multiplexing.This may increase context switch times and reduce the precision of the results.

\end{itemize}

\section{Conclusions} \label{sec:concAndFutureWork}
Detection based countermeasures for micro-architectural attacks are promising as they can adapt easily to a wide range of attacks. However, a major shortcoming is the unacceptably high false positive rate. Contemporary research attempts to address this limitation  on by deploying sophisticated analysis techniques. While this has found to only marginally improve false positives, the overheads and latencies are affected considerably due to complex techniques used. Unlike these approaches, the reactive design of \name facilitates lightweight analysis which allows falsely flagged threads to recover quickly with minimum overheads of less than 1\% while still accurately detecting attacks.
To the best of our knowledge, \name is the first reactive countermeasure for micro-architectural attacks in the Operating System scheduler and is therefore, invariant to system load. Additionally, it just adds around 15 instructions per context switch thus has a negligible impact on context switch latencies. It thus opens avenues to security aware OS scheduler designs.

\bibliographystyle{plain}
\bibliography{references}

\end{document}